\documentclass[aps,twocolumn,prb,showpacs,multicol,amsmath,amssymb]{revtex4}
\usepackage[dvips]{graphicx}

\newcommand{\be}{\begin{equation}}
\newcommand{\ee}{\end{equation}}

\newcommand{\bea}{\begin{eqnarray}}
\newcommand{\eea}{\end{eqnarray}}
\newcommand{\bd}{\begin{displaymath}}
\newcommand{\ed}{\end{displaymath}}
\newcommand{\ba}{\begin{array}}
\newcommand{\ea}{\end{array}}
\newcommand{\bi}{\begin{itemize}}
\newcommand{\ei}{\end{itemize}}
\newcommand{\bc}{\begin{center}}
\newcommand{\ec}{\end{center}}
\newcommand{\bfl}{\begin{flushleft}}
\newcommand{\efl}{\end{flushleft}}
\newcommand{\bfr}{\begin{flushright}}
\newcommand{\efr}{\end{flushright}}

\def\6{\partial}

\def\={\!\!\!&=&\!\!\!}
\def\+{\!\!\!&&\!\!\!+~}
\def\-{\!\!\!&&\!\!\!-~}

\begin{document}

\title[]{Angular resolved specific heat in iron-based superconductors: the case
 for nodeless extended $s$-wave gap}
\author{A.V.~Chubukov$^{1}$, I. Eremin$^{2}$}
 \affiliation{$^1$Department of Physics, University of Wisconsin-Madison, Madison,
Wisconsin 53706, USA \\
$^2$Institut f\"ur Theoretische Physik III, Ruhr-Universit\"at Bochum, D-44801 Bochum, Germany}
\begin{abstract}
We consider
the variation of the field-induced component  of the specific heat
$C({\bf H})$ with the direction of the applied field in $Fe-$pnictides within quasi-classical Doppler-shift approximation,  with special emphasis to recent experiments on  FeSe$_{0.4}$Te$_{0.6}$ [Zheng et al., arXiv:1004.2236]. We show that for extended $s-$wave gap with no nodes, $C({\bf H})$ has $\cos 4 \phi$
 component, where $\phi$ is the angle between ${\bf H}$ and
the direction between hole and electron Fermi surfaces. The maxima of $C({\bf H})$ are at $\pi/4$, $3\pi/4$, etc. if the applied field is smaller than
$H_0 \leq 1T$, and at $\phi =0, \pi/2$, etc. if the applied field is larger than $H_0$. The angle-dependence of $C({\bf H})$, the positions of the maxima, and the relative magnitude of the oscillating component are consistent with the experiments performed in the field of $9T >> H_0$. We show that the observed $\cos 4 \phi$ variation does not hold if the $s-$wave gap has accidental nodes along the two electron Fermi surfaces.
\end{abstract}

\date{\today}


\maketitle

The symmetry and the structure of the superconducting gap in $Fe-$based superconductors have been subjects of numerous experimental and theoretical papers in recent years~\cite{kamihara08,mazin,scalapino,chubukov,wang}.
There is a growing consensus among researchers that (i) the gap has an extended $s-$wave symmetry -- it belongs to a symmetric $A_{1g}$ representation of the $D_{4h}$ symmetry group of a square lattice and its average values along electron and hole Fermi surfaces (FS) are of opposite signs; (ii) that superconductivity originates from umklapp process in which  pairs of fermions hop between conduction and valence bands; and (iii) that the pair-hopping interaction is a combination of a screened Coulomb interaction and a magnetic interaction, mediated by antiferromagnetic spin fluctuations.

A more subtle and currently  hotly debated issue  is whether the gap has nodes. This is not a symmetry issue as, quite generally, extended $s-$wave gap can be approximated by a constant only along the hole FS, while
along the two electron FSs it has angle-independent and $\cos 2 \psi$ components: $\Delta_{e} (\psi) = \Delta_e (1 + b \cos 2 \psi)$, where $\psi$ is the angle counted from the line connecting the two FSs, and the sign of $\Delta_e$ is opposite to the sign of the gap along hole FSs.~\cite{comm} Such
$\Delta_e (\psi)$ has no nodes if $b<1$ and has ``accidental'' nodes  when $b >1$ at non-symmetry selected directions $\cos 2\psi = 1/b$
Because $Fe-$pnictides are multi-orbital systems, the $\cos 2\psi$ component of the interaction is generally not small,
i.e., $b$ can be either larger or smaller than 1, depending on the material.
Furthermore, $b$ gets larger when one includes into the gap equation
intra-band Coulomb repulsion because this term couples  to the gap averaged
over the FS and hence reduces angle-independent gap components but does not affect $\cos 2\psi$ components~(\onlinecite{chubukov}e). As a consequence,
$b$ becomes progressively larger as the system moves further away from the SDW phase and the effect of intra-band repulsion grows, that is,  overdoped ferropnictides are more likely to have nodes in the gap.

The issue whether or not the gap in $Fe$-pnictides has nodes is crucial for the understanding of low-energy properties of these materials and deserves a careful study.  The subject of this work is the interpretation of recent high-accuracy measurements~\cite{zheng} of the dependence of the low-temperature specific heat in FeSe$_{0.4}$Te$_{0.6}$ on the direction of a magnetic field.  Similar experiments have been carried out in the past on borocarbides~\cite{boro}, and heavy-fermion CeCoIn$_5$~(Ref. \onlinecite{before}) and revealed modulations generally
consistent with the $d$-wave gap (for details see \cite{anton,machida}; for experiments on thermal conductivity see~\cite{thermal}).

The generic reason for field-induced modulations of specific heat and thermal conductivity  in unconventional
superconductors is that a magnetic field induces vortices along the field direction.  In a vortex state of a type-II superconductor, scattering of quasiparticles on vortices  gives rise to a non-zero density of states (DOS) at zero energy. The magnitude of this residual DOS depends on the angle the field makes with the position of the minima of the modulus of the superconducting gap. This leads to modulations of the field-induced
linear-in-$T$ term in the specific heat~\cite{anton,machida,previous}. This reasoning works the best when the gap has nodes but should be generally applicable also to materials where the gap varies along the FS but not necessary has nodes, provided that the field is not too small.
Fe-pnictides are strong type-II superconductors (both magnetic and coherence lengths are of order $2-3\times 10^{2} \AA$, much smaller than the penetration depth $\lambda \sim 3 \times 10^3 \AA$~\cite{ruslan}), and vortex state extends to almost all fields [the upper critical field is about $100 T$,
lower critical field is about $H_{c1}\leq 10mT$\cite{lowcrit}].

\begin{figure}[t]
\centering
 \includegraphics[width=1.0\linewidth]{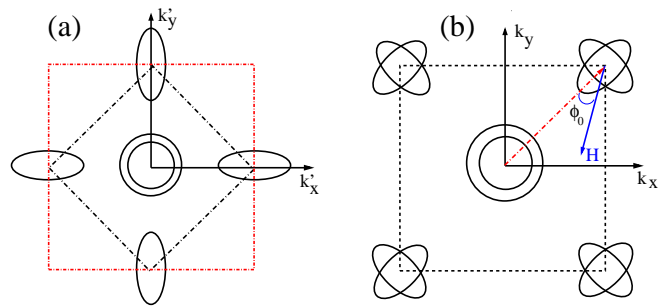}
\caption{(color online) Fermi surface topology of iron-based superconductors in the unfolded (a) and folded (b) BZ. There are two small hole pockets centered around the $\Gamma-$point and two elliptic electron pockets centered around the $(\pm \pi,0)$-point and $(0, \pm \pi)$-point of the unfolded BZ. Upon folding two electron pockets fold into the $(\pi,\pi)$-point of the folded BZ (dashed-dotted line in (a) is the boudary of the folded BZ). The magnetic field rotation is shown by angle $\phi_0$.}
\label{fig1}
\end{figure}

In  FeSe$_{0.4}$Te$_{0.6}$, the data~\cite{zheng} show~ $\cos 4 \phi$ modulation of $C({\bf H})$, with the magnitude of about $1\%$ of the total field-induced
$C({\bf H})$.
The maximum of  $C({\bf H})$ is at $\phi =0, \pi/2$, etc. what correspond to the
directions of ${\bf H}$ along the axis between hole and one of electron FSs in the unfolded Brillouine zone (BZ) [along the diagonals in the folded BZ].
The $\cos 4 \phi$ modulation of $C({\bf H})$ was originally
interpreted~\cite{zheng}  as evidence for the nodes in the gap. However, to
be consistent with the observed near-perfect $\cos 4 \phi$ form of  $C({\bf H})$, the nodes have to be located precisely at $45^\circ$ with respect to the $x$ axis, i.e., right at the crossing points of  two electron FSs in the folded BZ (see Fig. \ref{fig1}). This  is generally incompatible with the
``accidental'' nodes located at some arbitrary angles $\phi$.
The authors of Ref. \onlinecite{zheng}  argued that the data are inconsistent with  no-nodal extended $s-$wave gap and an extended $s-$wave gap with accidental nodes. To explain the data, they included spin-orbit coupling and argued that it creates nodes on electron FSs at exactly $45^\circ$, even if the gap was nodeless in the absence of spin-orbit interaction.~\cite{zheng}.

In this communication, we argue that the data of Ref. \onlinecite{zheng} can be  actually explained {\it quantitatively} in a conventional semi-classical Doppler-shift scenario for field-induced oscillations of $C({\bf H})$, but only
if one assumes that the gaps along the electron FSs have no nodes.  The nodeless gap in  FeSe$_{0.4}$Te$_{0.6}$ has been extracted from STM data~\cite{hanaguri},
and we argue that STM and angle-resolved specific heat measurements are consistent with each other.

Our reasoning is two-fold: First, as we said, the two electronic gaps generally have the
forms $\Delta_e = \Delta(1 \pm b \cos 2 \psi)$. The formula for the specific heat~\cite{anton} contains $\Delta^2_e$, i.e  $\cos 2\psi$ and $\cos 4 \psi$ terms.
 The $\cos 2\psi$ terms cancel out when the contributions from the two electron FSs are added, while the $\cos 4\psi$ term generates $\cos 4\phi$ modulation of  $C({\bf H})$. This, however, holds only if $\Delta_e$ does
 not have nodes ($b < 1$), otherwise the modulation of $|\Delta_e|$ will be more complex leading to a more complex structure of $C({\bf H})$.
Second, to be consistent with the data, the sign of $\cos 4\phi$ modulation
of  $C({\bf H})$ should be positive (maxima should be at $\phi=0, \pi/2$, etc).
At small ${\bf H}$ this is not the case -- the sign is negative.
 We show, however, that the sign of the $cos 4 \phi$ term in   $C({\bf H})$ depends on the magnitude of the field and changes from negative to positive as the field increases. We estimated the field where the sign changes and
 found that it is about $1T$ for all $b <1$, much smaller than $9T$, at which experiments have been performed. In other words, at the field of $9T$, $\cos 4\phi$ oscillations of  $C({\bf H})$ have maxima at
$\phi=0, \pi/2$, etc for arbitrary strong oscillating component of $\Delta_e$, as long as it remains nodeless (i.e., as long as $b <1$).
Furthermore, at $9T$ field, the magnitude of the oscillating part of  $C({\bf H})$ is around $1\%$ of the total specific heat, like in the data~\cite{zheng},
 and this number weakly depends on $b$ except for very small values, where it vanishes as $b^2$. The conclusion of our analysis is that the data on $C({\bf H})$ are quite consistent with the ``conventional'' theory of field-induced oscillations, provided that the gaps along electron FSs have moderate $\cos 2\psi$ oscillations and no nodes.
The range $b \approx 1$ is a gray area, and in the presence of some amount of disorder modulations of  $C({\bf H})$ may still look like $\cos 4\phi$ even when the gap has pairs of weakly spaced ``accidental'' nodes. Still, a more natural explanation of the data in  $FeSe_{0.4}Te_{0.6}$ is that the gap has no nodes.

The sign change of the prefactor for the oscillating  $\cos 4\phi$ component
 in the specific heat and thermal conductivity
is the well-known phenomenon for $d-$wave superconductors. The detailed theoretical study of the sign variation of the prefactor for the
$\cos 4\phi$ term with changing magnetic field and temperature has been performed by Vorontsov and Vekhter (VV)~\cite{anton} and by Hiragi et al~\cite{machida}.
VV recently performed numerical analysis of the angular dependence of $C({\bf H})$ in the ferropnictides~\cite{anton_new} and found the change from negative to positive prefactor of the $\cos 4\phi$ term with increasing field and temperature.  Our results are fully consistent with theirs and provide analytical reasoning for the sign change of the $\cos 4 \phi$ term in the iron-based superconductors.

Like we said, scattering of quasiparticles on vortices gives rise to a finite
DOS at zero energy what in turn gives rise to a linear in $T$ specific heat:
\begin{equation}
C({\bf H}) = A T N ({\bf H}),~ N({\bf H}) =
 \int_0^{2\pi} \frac{d\psi}{2\pi} \int_0^\infty d \omega   \sum_j N^j({\bf H},\psi)
\label{eq:1}
\end{equation}
where $A$ is the overall factor and $j$ is the band index.
The experiment~\cite{zheng} has been performed at the low $T \sim 2.6K$
when terms of higher order in $T$ are irrelevant.

We consider FS geometry consisting of two hole FSs centered at $\Gamma$ point and two electron FSs centered at $(0,\pi)$ and $(\pi,0)$ in the unfolded BZ, or at $(\pi,\pi)$ in the folded BZ (Fig. \ref{fig1}). The potential presence of the third hole FS is not essential for our analysis because oscillations of
$C({\bf H})$ come only from the two electron bands.
For simplicity, we assume that all bands are circular, {\it i.e.}, neglect ellipticity of electron bands.
We will follow Ref.\cite{anton} and
employ the formula for $N^j({\bf H},\psi)$ obtained by solving semi-classical Eilenberger equations for a given vortex lattice within Brandt-Pesch-Tewordt (BPT) approximation in which the dependence on the normal Green's function on the center of mass coordinate of a pair is replaced by an average over a unit cell of the vortex lattice.  Hiragi et al~\cite{machida} computed the DOS beyond BPT approximation and obtained, but the changes turned out to be quite small.
Suppose that ${\bf H}$ is applied in $Fe-Fe$ plane, at
an angle $\phi$ with respect to the $x-$axis (which in momentum space is
the direction between hole and electron bands).  In the area surrounding the vortex, the DOS at zero energy can be generally written as~\cite{anton,previous,zheng}
\begin{equation}
N^j({\bf H},\psi) = \frac{\alpha({\bf H},\psi)}{\sqrt{\alpha^2({\bf H},\psi) + \left(\Delta^j (\psi)\right)^2}},
\label{eq:2}
\end{equation}
where $\alpha({\bf H},\psi)$ is proportional to the component of the Fermi velocity normal to the field $v^\perp_F = v_F \sin (\psi-\phi)$: $\alpha({\bf H}, \phi)= \bar{\alpha} \sin(\psi-\phi)$, where
$\bar{\alpha}=c v_F/(2\sqrt{2}\Lambda)$, $\Lambda=\sqrt{\hbar c /(2  |e|B})$ is the magnetic length, and $c = O(1)$ is a numerical
factor which carries information about the geometry of the
vortex structure and the distance from the vortex core~\cite{zheng}.

\begin{figure}[t]
\centering
 \includegraphics[width=1.0\linewidth]{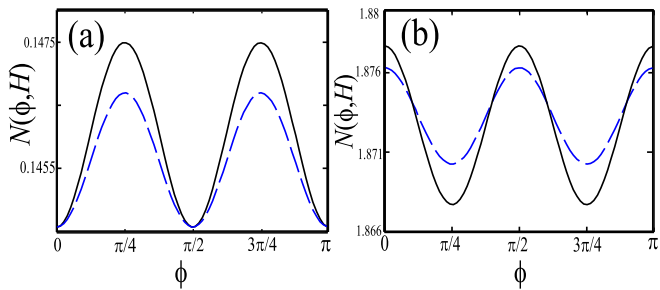}
\caption{ Calculated angular dependence of the density of states for
large $a =10$(a) and small $a=0.1$(b). We set $b =0.5$. The solid and dashed curves refer to the numerical solution of  Eq.(3) and the approximate analytical
formulas Eqs. (4) and (5), respectively. In panel (a) we matched analytical and numerical curves  at $\phi=0$.}
\label{fig2}
\end{figure}

Substituting $\Delta_e = \Delta (1 \pm b \cos 2 \psi)$ and $\Delta_h = {\text const}$  into (\ref{eq:2}) and shifting the integration variable, we obtain $N({\bf H})
= N_h(H) + N_e ({\bf H})$, where $N_h$ is independent on the direction of ${\bf H}$, and
$N_e ({\bf H}) = \int_{0}^{2\pi} \frac{d\psi}{2\pi} (N_e ({\bf H}, \psi) + N_e ({\bf H}, \psi + \pi/2)$, where
\begin{equation}
 N_e ({\bf H}, \psi) = \sqrt{\frac{1}{1+ \frac{a^2\left(1+ b \cos \left(2\psi + 2\phi\right)\right)^2}{ \sin^2\psi}}}
\end{equation}
and $a^2=\frac{\Delta^2}{\bar{\alpha}^2} = H_0/H$.

Consider the limits of small and large $a$ separately. At large $a$ (small fields) $N_e ({\bf H}) \propto 1/a$. Oscillating component of $N_e$ can be obtained analytically at small $b$. Expanding in $b$ we find
\be
N_e ({\bf H}) = N_e (H) - \frac{2 b^2}{15\pi |a|} \cos 4 \phi + O\left(b^4\right)
\label{4}
\ee
where $N_e (H)$ is a $b-$dependent non-oscillating term.
We see that the DOS does contain $\cos 4\phi$ oscillations, as we anticipated,
however the sign of the $\cos 4\phi$ term is negative, which implies that $C({\bf H})$ has peaks at $\pi/4, 3\pi/4$, etc, in disagreement with the data.
At larger $b$, the oscillating part of $N_e ({\bf H})$ contains higher harmonics $\cos 8 \phi, \cos 12 \phi...$ and has to be calculated numerically. We present the results in Fig. \ref{fig2}(a).  We see that  the oscillating part of
$N_e({\bf H})$  still well described by $\cos 4 \phi$ form for arbitrary $b <1$,
 despite that higher harmonics are not parametrically small. Also, the sign of the oscillating part remans negative at large $a$ for arbitrary $b <1$ (i.e., for all $b <1$, the maxima of $C({\bf H})$ are at $\pi/4$, etc.).

The situation changes, however, in the opposite limit of large fields, when $a <<1$. Now $N_e ({\bf H})$ can be expanded in $a$. The expansion requires care because of infra-red divergencies and yields, {\it at arbitrary $b <1$}
\be
N_e ({\bf H}) = 2 - \frac{4 a}{\pi} + 2 a^2 b^2 \cos{4 \phi}~ \left(1 + O(a)\right) +...
\label{5}
\ee
where dots stand for terms of order $a^4$, at which order higher harmonics appear.  We see that the oscillating component is now $\cos 4\phi$ for all $b <1$, and the sign of the oscillating part is positive, i.e., the maxima of
$C({\bf H})$  are now at $\phi =0, \pi/2...$, like in the experimental data~\cite{zheng}.
In Fig. \ref{fig2}(b) we present the result of numerical evaluation of $N_e ({\bf H})$ and compare it with Eq. (\ref{5}).  Clearly, there are $cos 4\phi$ oscillations with a positive prefactor.
\begin{figure}[t]
\centering
 \includegraphics[width=1.0\linewidth]{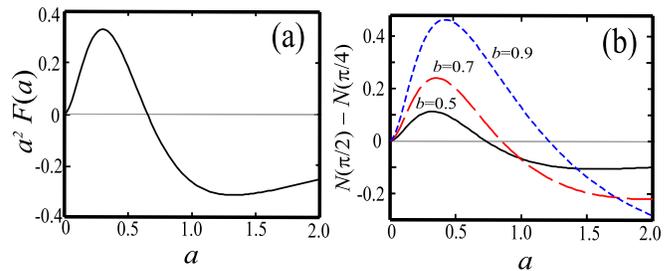}
\caption{(color online) (a) The functional form of the function
 $a^2F(a)$ from Eqs. (6) and (7). Sign change of $F(a)$ implies sign change of the prefactor for $\cos 4\phi$ term in the density of states; (b) the difference between $N(\phi)$ for $\phi=\pi/2$ and $\phi=\pi/4$ from Eq. (3)
as a function of $a$ for various $b$. For any $b$ from the interval
$0<b<1$, $N(\pi/2)-N(\pi/4)$ changes sign at a finite $a=a_0$, ranging between 0.65 and 2.8.}
\label{fig3}
\end{figure}
The value of $a$ at which the oscillating part of $N_e ({\bf H})$ changes sign,
and the crossover from a small field to a high field behavior can be analyzed analytically at small $b$. Expanding Eq.(3) in $b$ to order $b^2$ and integrating over $\psi$ we obtain
\be
N_e ({\bf H}) = \frac{4}{\pi} \arctan \frac{1}{a}  + a^2 b^2
\cos 4 \phi  F(a)
\label{6}
\ee
where
\be
F(a) = \frac{1}{2\pi}\int_0^\pi d\psi \frac{\sin{\psi} \cos{4\psi}}{(a^2 + \sin^2{\psi})^{5/2}} \left(2 a^2 - \sin^2{\psi}\right)
\label{7}
\ee
We plot $a^2F(a)$ in Fig. \ref{fig3}(a). This function changes sign at $a = a_0 \approx 0.65$ and is negative at larger $a$ (smaller fields).
 The implication of this result is that $C({\bf H})$ changes sign at a {\it finite} field even when the gap anisotropy is infinitesimally small.  We analyzed the evolution of $a_0$ with increasing $b$ and found [Fig. \ref{fig3}(b)] that $a_0$ remains finite and of order one for all $b <1$.  Observe also that $a^2F(a)$ is of order $10^{-2}$ for all $a$ except for the smallest one, i.e for $b \sim 1$, the oscillating part of $C({\bf H})$ is of order $10^{-2}$
 of the total  $C({\bf H})$.

\begin{figure}[t]
\centering
 \includegraphics[width=1.0\linewidth]{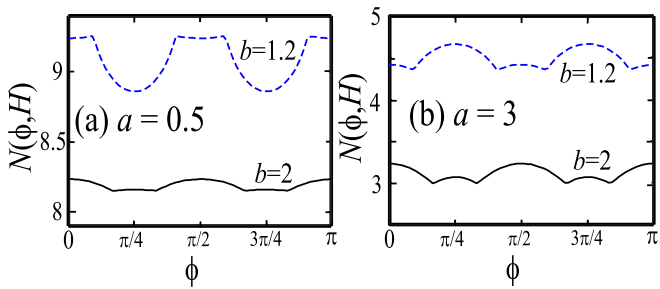}
\caption{The angular dependence of the density of states, Eq. (3),
 for $b>1$. The form of $N(\phi, H)$ is clearly different from $\cos 4 \phi$.
Additional minima or maxima correspond to the nodes at $\cos 2\psi =1/b$.
 Panel (a) -- large fields, $a<1$, panel (b) --  small fields,
$a>1$. The sign of the oscillating component still depends on whether or not
$a$ exceeds some $b-$dependent critical value.}
\label{fig4}
\end{figure}

In Fig. \ref{fig4} we present $N_e ({\bf H})$ for $b >1$, when the gap along electron FSs has accidental nodes at $\cos 2\psi =1/b$. We clearly see that the angular dependence is different from $\cos 4\phi$ -- there appear additional
 maxima or minima in  $N_e ({\bf H})$ associated with zeros of $\Delta_e$ (minima of $|\Delta_e|$). These deviations from $\cos 4 \phi$ form for $b >1$ have been reported before.~\cite{graser}

To estimate the value of $H_0$ in $a = H_0/H$, we use $\Lambda \approx
180 \AA/ \sqrt{B}$, where $B$ is the value of a field in Tesla, and take
$v_F \sim 3.3 \times 10^5 m/s$, averaged between two  electron bands~\cite{zheng}, and $\Delta_e \sim 4 meV$~\cite{Delta,park,homes}.  We obtain
$H_0 = (0.89/c^2) T$. Hanaguru et al~\cite{hanaguri} extracted a smaller $\Delta \sim 1.7 meV$ from their STM data. This will lead to even smaller $H_0 \sim (0.16/c^2)$.  The value of $c$ is not known but should generally be of order one. For $c \leq 1$, $H_0$  is well below $9T$ at which the experiments are performed. In other words, $H=9T$ is deep inside the range of $H > H_0$, when the oscillating part of $C({\bf H})$ has $\cos 4\phi$ form with the
maxima at $\phi =0, \pi/2$, etc.  Using $H_0 \sim 1T$
for definiteness and collecting the contributions to $N({\bf H})$ from two hole and two electron bands, we found that the amplitude of the oscillating part of $C({\bf H})$ is $0.028(2b)^2$ of the total  $C({\bf H})$, which for $2b = O(1)$ is quite consistent with one percent effect observed in the experiment.~\cite{zheng}

To conclude, we considered analytically, within BPT approximation,
 the variation of the field-induced component of the specific heat with the direction of the applied field. We demonstrated that this scenario yields the $\cos 4 \phi$ variation with the maxima at $\pi/4, 3\pi/4$, etc if the applied field is smaller than $H_0 \leq 1T$, and  $\cos 4 \phi$ variation with the maxima at $\phi =0, \pi/2$, etc if the applied field is larger than $H_0$. Both results
are valid provided that the gaps along electron FSs have $\cos 2\psi$ component, but no nodes.   We argued that
the $\cos 4 \phi$ form of oscillating part, the positions of the maxima, and the relative magnitude of the oscillating component of $C({\bf H})$ are consistent with the experiments by Zheng et al~\cite{zheng} performed in the  $9T$ field, well above $H_0$.
We therefore argue that the data on the angular dependence of field-induced $C({\bf H})$ in FeSe$_{0.4}$Te$_{0.6}$ are actually consistent with no-nodal extended $s-$wave gap in this material.
 The same no-nodal extended s-wave gap has been extracted from STM~\cite{hanaguri}, Andreev reflection~\cite{park},
 and optical conductivity~\cite{homes} data on  FeSe$_{0.45}$Te$_{0.55}$.

We thank
 C. Homes,
 I. Mazin,  H-H. Wen, and particularly A. Vorontsov  for useful conversations.
A.V.C. acknowledges the support from NSF-DMR 0906953.
I.E. acknowledges the support from  the RMES Program (Contract No. N
2.1.1/2985). A.V.C. is thankful to MPIPKS in Dresden for hospitality during the work on the manuscript.

\end{document}